# Advanced Accelerator Concepts: From Birth to High Impact Science

James Rosenzweig
*Department of Physics and Astronomy*
*University of California, Los Angeles*
Los Angeles, California, USA
rosen@physics.ucla.edu

*Abstract*—This recounting of the history of the last three-and a-half decades of advanced accelerator concepts is offered from a decidedly parochial point of view – that of the career of the author, Prof. James Rosenzweig of the UCLA Dept. of Physics and Astronomy. This short voyage through a by-now long career will illustrate the very beginning of the compelling field of advanced accelerators, proceed through their maturation into one of the fastest growing areas of beam-based science, and give a look into their emerging importance in applications. An important aspect of advanced accelerators is their relationship to other burgeoning fields, particularly free-electron lasers. The framework of this retelling lends itself particularly well to illustrating this relationship. Likewise, this quick summary serves to demonstrate the essential team nature of our field, and the contributions of participants from all levels, ranging from students to those scientists whose careers may have developed in previous eras of positive ferment in accelerator science.

## I. Introduction

This article has been composed by the author, Prof. James Rosenzweig of the UCLA Dept. of Physics and Astronomy, to give a written appreciation and historic context for his award of the 2022 Advanced Accelerator Prize. This honor has been given with the accompanying citation, which follows: "For his seminal and pioneering contributions at the nexus of advanced accelerators, light sources and beam physics". There is quite a bit to account for in this citation, and so the occasion presents itself to recount the story of the modern development of advanced, high field accelerators, based on emerging new techniques. This retelling begins very near to the start of the field, and progresses to the current time period, where advanced accelerators.

To tell this story in a familiar way, with the readers' indulgence, I will proceed to describe the my contributions and observations on the field directly in first person. I now appreciate, looking back, that my influence on the field of advanced accelerators has been extensive, ranging over a wide range of fields, from wakefields in plasmas and dielectrics, to laser-driven accelerators, and on to cryogenic, very-high-field accelerators. Indeed, with my vigorous research program and equally energetic educational efforts, my initiatives and accomplish-ments have marked not only the multitude of fields comprising advanced accelerator science, but also their burgeoning applications, particularly in the realm of new-generation light sources.

These nominally diverse fields have been brought together by the dedicated efforts of my group at UCLA, the Particle Beam Physics Laboratory (PBPL). The approach instituted at the PBPL, which is a type of "school" in the sense of intellectual flavor, has indeed produced a body of work that embraces a coherent view of advanced accelerators, fundamental beam physics, and new light sources. This viewpoint is shared by the large cohort of students and postdocs trained in the PBPL program at UCLA. This group has had a notable impact on the field, having been recognized with an impressive list of awards: (four) SLAC Panofsky Fellowships; (two) Young International FEL Prizes; the APS DPB Best PhD Thesis Award; CERN Marie Curie Fellowship; and the EPS Sacherer Prize for Young Accelerator Scientists, and the DOE HEP Early Career Award.

It is instructive to review the research achieve-ments motivating the recognition attendant to the 2022 AAC Prize, which have provided key direction to the development of the advanced accelerator concepts comm-unity for the last 35 years, or nearly the length of the field itself. I organize the list of these major accomplishments by theme, and include a discussion of present activity that follows on the previous work in the exciting field of advanced accelerator concepts. In order to properly set the stage, we begin with a brief discussion of my pre-UCLA years.

## II. Prehistory

Before the UCLA era began, my early career training occurred as a Wisconsin graduate student performing experimental and theoretical research at Argonne National Laboratory. The key players in advanced accelerator initiatives at Wisconsin set the tone for much of the subsequent work in the AAC field; they were Dave Cline, Fred Mills, and Sandro Ruggiero. Key support at the U.S. Dept. of Energy came from David Sutter, who was a keystone of the effort which gave birth to the field of advanced accelerators.

At the pioneering ANL laboratory termed the Advanced Accelerator Test Facility, or AATF, we performed the first proof-of-principle experiments on plasma wakefield acceleration [1], uncovering critically important focusing effects [2,3,4,5], and reporting the first aspects of nonlinear plasma waves [6]. This last work led to a deep investigation of nonlinear PWFA [7,8,9], and ultimately, at UCLA to my now- recognize as seminal proposal for operation of plasma wakefield accelerators in the highly nonlinear "blowout" regime, discussed below. This emphasis on nonlinear operation, which initially seemed to be a curious distraction in the PWFA field, indeed turned out to be essential.

During the time period of initial wakefield studies at ANL, I participated with some interest in the first demonstrations of dielectric wakefield accelerator (DWA) [10]. This also turned out to be a formative initiative, as I embraced the DWA line of inquiry two decades later, also similarly emphasizing a non-linear limit, where the properties of dielectrics begin to change due to field strengths at the gigavolt- per-meter level.

The scientists at the AATF that collaborated on both PWFA and DWA included Jim Simpson (group leader), Sekazi Mtingwa, Paul Schoessow, Wei Gai, and Sandro Ruggiero.

After the AATF, I accepted a Wilson Fellowship at Fermilab, one of the few instances for which this position was granted for a researcher from accelerator physics. At Fermilab, I concentrated on learning about colliders, and played a strong role in initiating the first wave of research on superconducting linear colliders [11], through comp-rehensive design studies. This initiative, initially aimed at the TESLA linear collider project, eventually morphed into the strong Fermilab effort towards realizing the International Linear Collider (ILC). A key collaborator in the effort at Fermilab was Helen Edwards.

It should be noted that this formative period is by now thirty years past, and the majority of the protagonists from that era are no longer living. Their legacy is another matter; it lives on quite vigorously.

## III. UCLA PWFA Research

Upon arriving at UCLA, I initially placed my emphasis on continuing superconducting linear collider work, concentrating on linear collider-quality RF photoinjectors [12]. This direction was natural, as at the time I began my time at UCLA collaborating experimentally on photoinjector development for the budding free-electron laser program at UCLA.

The situation where I de-emphasized PWFA research lasted only a few months, until met a talented computational plasma physicist named Boris Breizman, who gave me his program which was capable of simulating with minimal computational resources, the nonlinear response of plasmas to beams that were denser than the ambient plasma electron distribution. In this way I uncovered the PWFA blowout regime. This initiative [13] has had a profound effect on the subsequent dramatic progress in the PWFA field, as it identified new, highly advantageous properties in the plasma "bubble" produced: linear focusing, and acceleration dependent only on wave phase. Thus, it has been possible to predict excellent beam phase space preservation, attributes that have since been verified experimentally.  has produced notable theoretical and experimental work exploring the PWFA in this nonlinear regime.

My own experimental work in the PWFA blowout regime has ranged from path breaking demonstrations of ion-focusing transport [14,15] as well as thin lens focusing [16], to first experiments on acceleration in this scheme [17,18]. I, along with many collaborators, investigated some of the first high impact proposals for injection into PWFAs, first using density transitions [19,20,21,22], a technique that endures today. Most recently, I co-led with Bernhard Hidding the successful first demonstration of injection of low emittance electron beams through laser-induced ionization, in the E210 Trojan Horse experiment at FACET [23,24].

These experiments have been accompanied by a wide range of theoretical studies. One particularly impactful thrust involved more controlled injection schemes, which give beams with the highly increased brightness one expects from scaling up the field and frequency of the wave capturing the beam [25]. This focus most notably included the original and continuing Trojan Horse theoretical treatments with Hidding [26,27,28,29,30,31]. Another area of high impact has been concentrated on the first predictions of ion collapse in the PWFA [32,33], which has now led to an experiment proposal at FACET-II [34]. Another fundamental subject has been investigated in detailed, that of the physics involved in the scaling of very nonlinear plasma wakes [35,36], which explains the persistence of linear Cerenkov scaling even in extremely nonlinear interactions.

PBPL PWFA research has continued in recent times to include experimental examinations of optimized transformer ratios [37] based on spatially shaped beams, as well associated diagnostic schemes [38]. New schemes for beam injection that are offshoots of the Trojan Horse "plasma photocathode" approach are being investigated at FACET-II. Finally, acceleration the TeV/m level in very dense plasmas, based on ultra-short beams or resonant excitation is being pursued [39].

## IV. UCLA DWA Research

In the past decade and one-half, the PBPL has led the effort to extend DWA to the GV/m frontier [40] and beyond, with first experiments at the SLAC FFTB showing over 5 GV/m longitudinal field before breakdown. These experiments led to follow on measurements of coherent Cerenkov radiation production from DWA [41] at UCLA Neptune Lab.

When the FFTB was replaced by FACET, there the PBPL program in DWA demonstrated sustained acceleration in structures up to 15 cm in length [42]. In this period of active DWA research at both FACET and the BNL ATF, new approaches to structure design were explored, including photonic concepts [43], and exploitation of new symmetries [44], to control mode content in 3D and to suppress beam breakup (BBU) [45,46]. As these are highly coupled structures, BBU is indeed thought to be a serious limitation of the DWA in application.

Dielectric wakefield studies at GeV/m gradients have uncovered unexpected new physics effects in the DWA interaction, such as high-field-induced conductivity [47]. Recently the PBPL team have investigated acceleration of positrons in DWA [48]. This work, along with the above-described PWFA research, has had an outsized effect on the programs at SLAC FACET/FACET-II, the Argonne Wakefield Accelerator and the BNL ATF.

These programs remain very active, with new experiments in DWA underway or planned, including a detailed understanding of normal and skew quadrupole effects in slab-symmetric structures. These strong quadrupole of focusing effects can be harnessed to provide new mechanisms of BNS damping, permitting meter-scale and GeV energy gain experimental scenarios at FACET-II [49].

In recent years, wakefield acceleration at the PBPL has increasingly relied on the leading contributions of UCLA scientist Gerard Andonian.

## V. LASER ACCELERATION RESEARCH

I became interested in laser-based acceleration schemes in the mid-1990's and published several seminal papers that have strongly guided the development of the dielectric laser accelerator, or DLA [50,51]. I led the GALAXIE DLA-based compact free-electron effort [52] for DARPA, which produced a deep understanding of the unique beam stability conditions in the DLA [53]. This program gave way to the highly successful ACHIP collaboration funded by the Moore Foundation. In this context at UCLA, Prof. Pietro Musumeci has been a key ACHIP protagonist

Laser-based acceleration at UCLA has another emphasis, also mainly led by Musumeci; that of the inverse free-electron laser (IFEL). I have had the pleasure of collaborating with him on several initiatives, including the first demonstration of very high energy gain in the IFEL [54], and production of high-quality accelerated beams [55]. This work culminated on with a joint project, mentioned in appropriate context below, in which an IFEL-derived beam was used in production of beam-laser scattered Compton X-rays.

## VI. ADVANCED ELECTRON SOURCES

Research on advanced electron sources has been critical in the development of the advanced accelerator and radiation production fields. As such, I have occupied myself with active research in this area. My resulting contributions range from the theoretical and experimental basis of high field particle dynamics [56] and emittance compensation [57,58,59], to the first exploration of the longitudinal blowout regime [60,61,62].

In this field, I have introduced a number generations of high brightness RF photoinjectors [63], which have been enabling technology for photoinjector laboratories worldwide (SLAC, BNL, UCLA, INFN-LNF Frascati, Sincrotrone Trieste, FNAL, LLNL). This program has included not only standard 1.6 cell RF guns, but new integrated systems, including the hybrid photoinjector [64,65,66,67]. Most recently, in collaboration with S. Tantawi and others, the PBPL has been developing cryogenic copper structures that we have collaboratively demonstrated support surface fields up to 500 MV/m [68,69].

This is an enabling technology for applications in high energy physics (the $C^3$ linear collider [70]) and free-electron lasers. At UCLA, in collaboration with SLAC, the first step in developing this approach is to base a new generation of very high brightness RF photoinjector operated at 250 MV/m peak field [71,72]. This device is capable of producing linear-collider-class asymmetric emittance beams when magnetized, and to drive new types of FELs (see below) when operated in high brightness mode.

## VII. ELECTRON BEAM MANIPULATION AND DIAGNOSIS

The PBPL program has played a pioneering role in the development of new methods for manipulating and diagnosing electron beams [73], so that new capabilities in advanced accelerator and light source research may be reached. The program has been a leader in the manipulation of high brightness beams, introducing or exploring new compression techniques based on chicanes [74,75,76], IFEL bunching [77], dogleg transport [78] and velocity bunching [79,80]) and beam shaping methods [81].

In this context we have necessarily introduced numerous influential measurement methods, including emittance diagnosis in the presence of space-charge [82], and a variety of coherent radiation-based [83,84,85,86,87] longitudinal diagnostics – addressing measurement challenges from the picosecond down to attosecond level.

## VIII. ADVANCED LIGHT SOURCES

Much of my work on electron sources was motivated by the needs of the first proof-of-principle self-amplified spontaneous emission free-electron laser (SASE FEL) experiments [88,89,90]. For the development of RF photoinjectors, and key participation (in the collaboration headed by Prof. Claudio Pellegrini) in the initial SASE FEL experiments I was awarded along with Ilan Ben-Zvi, the 2007 International Free-Electron Laser Prize. Special recognition was given for the introduction of start-to-end simulations to ascertain the microscopic physics of the beam-FEL interaction.

I have in the past decade merged a considerable FEL research effort (including orbital investigation angular momentum effects [91,92,93,94]) with advanced accelerators to produce a new concept – the *5th generation light source*. The PBPL is now working on several manifestations of this new class of instrument, including: MEMS-based undulators [95] driven by DLAs; inverse FEL acceleration to produce Compton X-rays [96]; and demonstrator FELs based on plasma accelerators (with INFN-LNF through the EuPRAXIA initiative [97], and with the LBNL BELLA team [98]). Our work on the Compton sources is the culmination of a steady campaign [99,100,101,102,103,104] to advance the physics of ICS sources.

The major focus of PBPL efforts on next-generation XFELs based on advanced accelerator methods, however, concentrates on cryo-RF at high field. This proposal, recently published in a highly influential article in *New Journal of Physics* [105], shows that the very high brightness beam produced by the new cryo-gun, accelerated in >100 MeV/m cryo-linacs, paired with innovative compression methods, state-of-the-art undulators, and compact X-ray optics [106,107,108] can produce an extremely attractive XFEL. This instrument, the ultra-compact XFEL (UC-XFEL), is aimed at revolutionizing access to XFEL facilities; its compact size (<40 m) and modest cost (~$35M) should permit it to be diffused widely in university or industry labs. The push towards UC-XFEL is also recognized as highly synergistic with the needs to beam physics and technology development for $C^3$.

## IX. NEW DIRECTIONS

Extending interest in high gradient-based electron source, of late the PBPL and collaborators (notably Peter Hommelhoff, Univ. of Erlangen) have utilized very high field (to 30 GV/m) laser-surface interactions in a nano-blade geometry to show extremely low emittance, femtosecond electron pulses that may be ideal injectors for DLAs, or be extended to linear collider asymmetric emittance sources [109,110]. The fields in this interaction are the largest non-destructive fields measured in such a device.

Together with DWA and cryo-RF initiatives, this illustrates an emerging theme in PBPL research, that of high field effects in solid-state matter. The nano-blade initiative, as well as that of cryogenic gun and the UC-XFEL, are vital components of the successful NSF STC, the Center for High Brightness Beams, that the PBPL has played a strong role in for the last six years.

In addition to his vigorous programs at national user facilities, we have now constructed a new, ambitious laboratory at UCLA, heir to previous efforts (this is the fourth photoinjector lab constructed by the PBPL at UCLA), that will be a venue for wakefield acceleration, compact light (FEL and ICS) sources, and frontier high brightness beam sources. This lab will host the frontier cryo-RF development at UCLA, and its proximity to the Basic Plasma Science Facility will permit new wakefield studies that explore phenomena related to long time-scale PWFA behavior, in the laboratory and in the space environment [111].

## X. SUMMARY

The contributions described above have produced several tangible lasting features, prominent among them; and the training of a very large cohort of graduate students who have had major secondary influence in the field of advanced accelerators. In addition, my laboratory program has produced 15 patents applied for and/or granted, and spun off a highly successful company, RadiaBeam Tech-nologies, which has played a key role in development of beam science, technology and advanced accelerator methods. This contribution has led to a notable strengthening of the US industrial accelerator landscape, and gave further dimension to the impact of PBPL trainees.

It is only appropriate that we list here these trainees, those who spent all or part of their graduate or post-doctoral career at the UCLA PBPL training for a career in advanced accelerators and related fields:

- Gil Travish
- Nikolai Barov
- Eric Colby
- Aaron Tremaine
- Hyyong Suk
- Matthew Thompson
- Salime Boucher
- Ron Agusstson
- Adnan Doyuran
- Alex Murokh
- Oliver Williams
- Kip Bishopfberger
- Gerard Andonian
- Pedro Frigola
- Scott Anderson
- Rodney Yoder
- Luigi Faillace
- Alan Cook
- Joel England
- Yusuke Sakai
- Atsushi Fukasawa
- Alessandra Valloni
- Agostino Marinelli
- Gabriel Marcus
- Erik Hemsing
- Andrey Knayzik
- Josh McNeur
- Diktys Stratakis
- Sam Barber
- Aihua Deng
- Yunfeng Xi
- Alex Cahill
- Brendan O'Shea
- Finn O'Shea
- Claudio Emma
- Egor Dyunin
- Ariel Nause
- Phuc Hoang
- Ryan Roussel
- Ivan Gadjev
- Nathan Majernik
- Joshua Mann
- Gerard Lawler
- Pratik Manwani
- Monika Yadav
- Walter Lynn
- Fabio Bosco

I am truly grateful to have had the opportunity to provide mentorship to these colleagues.

## ACKNOWLEDGMENT

I would like to thank the following agencies for their support over the course of my career: US Dept. of Energy High Energy Physics and Basic Energy Sciences; US National Science Foundation; Defense Advanced Research Projects Agency; US Domestic Nuclear Detection Office; Italian Istituto Nazionale di Fisica Nucleare; Israel Ministry of Defense; the Keck Foundation; the Sloan Foundation.

## REFERENCES

[1] "Experimental Observation of Plasma Wake-field Acceleration", J.B. Rosenzweig, *et al.*, *Phys. Rev. Letterss* **61**, 98 (1988).
[2] Beam Optics of a Self-focusing Plasma Lens", J.B. Rosenzweig and P. Chen, *Phys. Rev. D* **39**, 2039 (1989)

[3] "Final Focusing and Enhanced Disruption from an Underdense Plasma Lens in a Linear Collider", P. Chen, S. Rajagopalan, and J. Rosenzweig, *Phys. Rev. D* **40**, 923 (1989).
[4] "Experimental Studies of Plasma Wake-field Acceleration and Focusing", J. B. Rosenzweig, *et al.*, *Physica Scripta* **T30**, 110 (1990).
[5] "Demonstration of Electron Beam Self-focusing in Plasma Wake-fields", J.B. Rosenzweig, *et al,*, *Phys. Fluids B* **2**, 1376 (1990).
[6] "Experimental Measurement of Nonlinear Plasma Wake-fields", J. B. Rosenzweig, *et al.*, *Phys. Rev. A - Rapid Comm.* **39**, 1586 (1989
[7] "Nonlinear Plasma Dynamics in the Plasma Wake-Field Accelerator", J.B. Rosenzweig, *Phys. Rev. Lett.* **58**, 555 (1987)
[8] "Trapping, Thermal Effects and Wave Breaking in the Nonlinear Plasma Wake-field Accelerator", J.B. Rosenzweig, *Phys.Rev. A* **38**, 3634 (1988).
[9] "Multi-Fluid Models for Plasma Wake-field Phenomena" J. B. Rosenzweig, *Phys. Rev. A* **40**, 5249 (1989).
[10] "Experimental Demonstration of Wake-field Effects in Dielectrics", W. Gai, *et al.*, *Phys. Rev. Lett.* **61**, 2756 (1988)
[11] "Parameter Lists for TESLA", J.B. Rosenzweig, in *Proceedings of the 1st International Workshop on TESLA -- a TeV Superconducting Linear Collider'*, H. Padamsee, Ed., 180 (Cornell, 1990).
[12] "Flat-Beam RF Photocathode Sources for Linear Collider Applications", J.B. Rosenzweig, in the *Proceedings of the 1991 IEEE Particle Accelerator Conference* **1**, 1987 (IEEE, New York, 1991).
[13] "Acceleration and Focusing of Electrons in Two-Dimensional Nonlinear Plasma Wake-fields", J. B. Rosenzweig, *et al.*, *Phys. Rev. A -- Rapid Comm .* **44**, R6189 (1991).
[14] "Propagation of Short Electron Pulses in an Underdense Plasma", N. Barov and J.B. Rosenzweig, *Phys. Rev. E* **49** 4407 (1994).
[15] "Propagation of Short Electron Pulses in a Plasma Ion Chanel" N. Barov, M.E. Conde, W. Gai, and J.B. Rosenzweig, *Physical Review Letters.* **80**, 81 (1998)
[16] "Propagation of Short Electron Pulses in a Plasma Ion Chanel" N. Barov, M.E. Conde, W. Gai, and J.B. Rosenzweig, *Physical Review Letters.* **80**, 81 (1998)
[17] "Observation of plasma wakefield acceleration in the underdense regime" N. Barov, J. B. Rosenzweig, M. E. Conde, W. Gai, and J. G. Power *Physical Review Special Topics Accelerators and Beams* **3** 011301 (2000).
[18] "Ultra High-Gradient Energy Loss by a Pulsed Electron Beam in a Plasma", N. Barov, *et al.,* in Proc. of the 2001 Particle Accel. Conference, 126 (IEEE, 2001)
[19] "Plasma Electron Trapping and Acceleration in a Plasma Wake Field Using a Density Transition", H. Suk, N. Barov, J. B. Rosenzweig, E. Esarey, *Phys. Rev. Lett.* **86**, 1011 (2001).
[20] "Plasma electron fluid motion and wave breaking near a density transition", R. J. England, J. B. Rosenzweig, and N. Barov *Phys. Rev. E* **66**, 016501 (2002).
[21] "Plasma density transition trapping as a possible high-brightness electron beam source" M. C. Thompson, J. B. Rosenzweig, and H. Suk, *Phys. Rev. ST Accel. Beams* 7, 011301 (2004).
[22] "Plasma electron fluid motion and wave breaking near a density transition", R. J. England, J. B. Rosenzweig, and N. Barov *Phys. Rev. E* **66**, 016501 (2002).
[23] A. Deng, et al. , "Generation and acceleration of electron bunches from a plasma photocathode", *Nature Physics* **15**, 1156–1160 (2019).
[24] D Ullmann, et al., "All-optical density downramp injection in electron-driven plasma wakefield accelerators" *Physical Review Research* **3** 043163 (2021)
[25] "Charge and Wavelength Scaling of RF Photoinjector Designs", J.B. Rosenzweig and E. Colby, *Advanced Accelerator Concepts* p. 724 (AIP Conf. Proc. 335, 1995).
[26] "Trojan Horse Laser Electron Injection and Acceleration in a Beam-Driven Plasma Blowout" B. Hidding, J. B. Rosenzweig, T. Konigstein, D. Schiller, D. L. Bruhwiler", *Phys. Rev. Lett.* **108,** 035001 (2012)
[27] "Hybrid modeling of relativistic underdense plasma photocathode injector", Y. Xi, B. Hidding, D. Bruhwiler G. Pretzler, J.B. Rosenzweig, *Physical Review ST-Accel. Beams,* 16, 031303 (2013)
[28] G. Wittig, et al., "Optical plasma torch electron bunch generation in plasma wakefield accelerators", *Phys. Rev. ST Accel. Beams* **18**, 081304 (2015).
[29] "Hot spots and dark current in advanced plasma wakefield accelerators", *Phys. Rev. ST Accel. Beams* **19**, 011303 (2016)
[30] G. G. Manahan, et al., "Single-stage plasma-based correlated energy spread compensation for ultrahigh 6D brightness electron beams" *Nature Communications* 8, 15705 doi:10.1038/ncomms15705 (2017)
[31] "Advanced schemes for underdense plasma photocathode wakefield accelerators: pathways towards ultrahigh brightness electron beams", *Royal Society Phil. Trans. A* (2019)
[32] "The Effects of Ion Motion in Intense Beam-Driven Plasma Wakefield Accelerators" J.B. Rosenzweig, A.M. Cook, A. Scott, M.C. Thompson, R. Yoder, *Phys. Rev. Lett* **95,** 195002 (2005)
[33] Claire Hansel, Monika Yadav, Pratik Manwani, Weiming An, Warren Mori, James Rosenzweig "Plasma Wakefield Accelerators with Ion Motion and the E-314 Experiment at FACET-II" arXiv:2107.00054 (2021)
[34] "Plasma Wakefield Accelerators with Ion Motion and the E-314 Experiment at FACET-II", Claire Hansel, Monika Yadav, Pratik Manwani, Weiming An, Warren Mori, James Rosenzweig arXiv:2107.00054 (2021)
[35] "Energy loss of a high charge bunched electron beam in plasma: Simulations, scaling, and accelerating wakefields" J. B. Rosenzweig, et al., *Phys. Rev. ST Accel. Beams* 7, 061302 (2004)
[36] "Energy loss of a high charge bunched electron beam in plasma: Analysis" N. Barov, J. B. Rosenzweig, M. C. Thompson, and R. B. Yoder, *Phys. Rev. ST Accel. Beams* **7**, 061301 (2004)

[37] R. Roussel, et al., "Single Shot Characterization of High Transformer Ratio Wakefields in Nonlinear Plasma Acceleration", *Phys. Rev. Lett.* **124**, 044802 (2020)

[38] "Longitudinal current profile reconstruction from a wakefield response in plasmas and structures", R. Roussel, G. Andonian, J. B. Rosenzweig, and S. S. Baturin, *Phys. Rev. Accel. Beams* 23, 121303 (2020)

[39] P. Manwani, N. Majernik, M. Yadav, C. Hansel, and J. B. Rosenzweig "Resonant excitation in plasma wakefield accelerators by optical-period bunch trains" *Phys. Rev. Accel. Beams* **24**, 051302 (2021)

[40] "Breakdown limits on Gigavolt-per-Meter Dielectric Wakefields", M.C. Thompson, et al., *Phys. Rev. Lett.,* **100**, 21 (2008)

[41] "Observation of Narrow-Band Terahertz Coherent Cherenkov Radiation from a Cylindrical Dielectric-Lined Waveguide," A. M. Cook, R. Tikhoplav, S. Y. Tochitsky, G. Travish, O. B. Williams, and J. B. Rosenzweig, *Physical Review Letters* **103**, 095003 (2009)

[42] B.D. O'Shea, et al., "Demonstration of Gigavolt-per-mete Accelerating Gradients using Cylindrical Dielectric-lined Waveguides" *Nature Communications* 12763 (2016).

[43] P. D. Hoang, et al., "Observation of wakefields from three-dimensional dielectric-based photonic structures", *Phys. Rev. Lett.* 120, 164801 (2018)

[44] "Electromagnetic Wake-fields in Slab-symmetric Dielectric Structures" A. Tremaine, J. Rosenzweig, P. Schoessow, *Physical Review E* **56**, 7204 (1997)

[45] "Dielectric wakefield acceleration of a relativistic electron beam in a slab symmetric structure", G. Andonian, et al., *Phys. Rev. Letters* **108**, 244801 (2012)

[46] B. D. O'Shea, et al., "Suppression of deflecting forces in planar-symmetric dielectric wakefield accelerating structures with elliptical bunches" *Phys. Rev. Lett.* 124, 104801 (2020)

[47] B. D. O'Shea, et al., "Conductivity induced by high-field terahertz waves in dielectric material", *Phys. Rev. Lett.* 123, 134801 (2019)

[48] N Majernik, et al., "Positron driven high-field terahertz waves via dielectric wakefield interaction'" *Physical Review Research* **4** 023065 (2022)

[49] J. B. Rosenzweig "Physics Goals of DWA Experiments at FACET-II", Proc. 2021 Int. Particle Accel. Conference (JACOW, 2021).

[50] A Proposed Dielectric-loaded Resonant Laser Accelerator", J.B. Rosenzweig, A. Murokh, and C. Pellegrini, *Phys. Rev. Letters* **74**, 2467 (1995).

[51] R.J. England, et al., "Dielectric Laser Accelerators", *Rev. Mod. Phys.* 86, 1337 (2014)

[52] "The GALAXIE All-Optical FEL Project", J. B. Rosenzweig, et al., *Proceedings of the 2012 Advanced Accelerator Concepts Workshop*, AIP Conf. Proc. 1507, 493 (AIP, 2012).

[53] "Stable charged particle acceleration and focusing in a laser accelerator using higher spatial harmonic resonance," B. Naranjo, A. Valloni, S. Putterman and J.B. Rosenzweig, *Physical Review Letters* **109**, 164803 (2012)

[54] "High Energy Gain of Trapped Electrons in a Tapered, Diffraction-Dominated In-verse-Free-Electron Laser" P. Musumeci, et al., *Phys. Rev. Lett.* **94**, 154801 (2005)

[55] "High Quality Electron Beams from Helical Inverse Free-Electron Laser Accelerator", J. Duris, et al., *Nature Comm.* 5, 4928

[56] "Transverse Particle Motion In Radio-Frequency Linear Accelerators", J.B. Rosenzweig and L. Serafini, *Physical Review E* **49**, 1499 (1994).

[57] "Envelope Analysis of Intense Relativistic Quasi-Laminar Beams in RF Photoinjectors: A Theory of Emittance Compensation", Luca Serafini and James Rosenzweig *Physical Review E* **55**, 7565 (1997).

[58] "Direct measurement of the double emittance minimum in the beam dynamics of the SPARC high-brightness photoinjector," M. Ferrario, et al., *Phys. Rev. Lett*. 99, 234801 (2007).

[59] Nonequilibrium transverse motion and emittance growth in ultra-relativistic space-charge dominated beams", S.G. Anderson and J.B. Rosenzweig, *Physical Review Special Topics - Accelerators and Beams* **3**, 094201 (2000

[60] "Longitudinal phase space characterization of the blow-out regime of rf photoinjector operation," J. T. Moody, P. Musumeci, M. S. Gutierrez, J. B. Rosenzweig, and C. M. Scoby, *Phys. Rev. ST Accel. Beams* 12, 070704 (2009)

[61] "Measurement of self-shaped ellipsoidal bunches from a photoinjector with post-acceleration", B.D. O'Shea, J. B. Rosenzweig, G. Asova, J. Bahr, M. Hanel, Y. Ivanisenko, M. Khojoyan, M. Krasilnikov, L. Staykov, and F. Stephan .*Phys. Rev. ST-Accel. Beams*. **14,** 012801 (2011)

[62] "Experimental Generation and Characterization of Uniformly Filled Ellipsoidal Electron-Beam Distributions", P. Musumeci, J.T. Moody, R.J. England, J.B. Rosenzweig, and T. Tran, *Phys. Rev. Letter.* **100,** 244801 (2008)

[63] "The High Brightness Electron Beam Physics and Photoinjector Technology Program at UCLA", J.B. Rosenzweig, in the *46th ICFA Beam Dynamics Newsletter*, Ed. Miguel Furman, pp. 119-150 (IUPAP, August 2008)

[64] "Design and applications of an X-band hybrid photoinjector", J.B. Rosenzweig, *et al.,* Nuclear Instruments and Methods A **657,** 107 (2011).

[65] A. Fukasawa, et al., 'Progress on the Hybrid Gun Project at UCLA" Physics Procedia: 52, 2-6 (2014)

[66] A. Nause, et al.,"6 MeV novel hybrid (standing wave - traveling wave) photo-cathode electron gun for a THz superradiant FEL", *Nuclear Instrum. Methods A* 1010 (2021)

[67] L Faillace, et al., "High field hybrid photoinjector electron source for advanced light source applications" *Physical Review Accelerators and Beams* 25 063401 (2022)

[68] A. D. Cahill, J. B. Rosenzweig, V. A. Dolgashev, S.G. Tantawi and S. Weathersby, "High Gradient Experiments with X-Band Cryogenic Copper Accelerating Cavities", *Physical Review Accel. Beams* 21, 102002 (2018)

[69] A.D. Cahill, J.B. Rosenzweig, V.A.Dolgashev, Z.Li, S.G. Tantawi, S. Weathersby, "rf losses in a high gradient cryogenic copper cavity" *Phys. Rev. Accel. Beams* **21,** 061301 (2018)

[70] Mei Bai, et al., "C3: A "Cool" Route to the Higgs Boson and Beyond" arXiv:2110.15800 (2021)

[71] J.B. Rosenzweig, et al., "Next Generation High Brightness Electron Beams from Ultra-High Field Cryogenic Radiofrequency Photocathode Sources", *Physical Review Accel. Beams* 22, 02340 (2019)

[72] R. Robles, Obed Camacho, Claire Hansel, Atsushi Fukasawa, Nathan Majernik, and J. B. Rosenzweig "Versatile, High Brightness, Cryogenic Photoinjector Electron Source" Phys. Rev. Accel. Beams 24, 063401(2021)

[73] "High brightness electron beam emittance evolution measurements in an rf photoinjector", A. Cianchi, et al. , *Phys. Rev. ST Accel. Beams,* **11,** 032801 (2008)

[74] "Pulse Compression of RF Photoinjector Beams: Advanced Accelerator Applications", J.B. Rosenzweig, N. Barov and E. Colby, *IEEE Trans. Plasma Sci.* **24,** 409 (1996)

[75] "Horizontal Phase Space Distortions Arising from Magnetic Pulse Compression of an Intense, Relativistic Electron Beam", S.G. Anderson, J.B. Rosenzweig, P. Musumeci, M.C. Thompson, *Phys. Rev. Lett*. 91, 074803 (2003)

[76] "Experimental characterization of the transverse phase space of a 60-MeV electron beam through a compressor chicane" F. Zhou, et al., *Phys. Rev. ST Accel. Beams* 9, 114201 (2006)

[77] River R. Robles and J.B. Rosenzweig "Compression of Ultra-High Brightness Beams for a Compact X-Ray Free-Electron Laser" *Instruments* 3(4), 53; (2019).

[78] "Sextupole correction of the longitudinal transport of relativistic beams in dispersionless translating sections", R. J. England, J. B. Rosenzweig, et al., R. Yoder *Physical Review ST-Accel. Beams* 8, 012801 (2005).

[79] "Velocity bunching of high-brightness electron beams" S. G. Anderson, et al., *Phys. Rev. ST-Accel. Beams* **8**, 014401 (2005).

[80] "Experimental Demonstration of Emittance Compensation with Velocity Bunching" M. Ferrario et al., *Phys. Rev. Lett*. **104,** 054801 (2010)

[81] "Generation and Measurement of Relativistic Electron Bunches Characterized by a Linearly Ramped Current Profile", R.J. England, J.B. Rosenzweig, and G. Travish, *Phys. Rev. Letter,* **100,** 214802 (2008)

[82] "Space-charge effects in high brightness electron beam emittance measurements", S.G. Anderson, J.B. Rosenzweig, G. P. Le Sage, J.K. Crane, Phys. Rev. ST Accel Beams **4**, 014201 (2001).

[83] "Coherent Transition Radiation Diagnosis of Electron Beam Microbunching", J. Rosenzweig, G. Travish and A. Tremaine, *Nucl. Instr. Methods A* **365** 255 (1995).

[84] "Bunch length measurement of picosecond electron beam from a photoinjector using coherent transition radiation" A. Murokh, J. Rosenzweig, *et al.*, *Nuclear Instruments and Methods A* **410,** 549 (1998).

[85] "Observation of Self-Amplified Spontaneous Emission-induced Electron Beam Microbunching Using Coherent Transition Radiation", A. Tremaine, J. B. Rosenzweig, *et al., Physical Review Letters,* **81** 5816 (1999).

[86] "Coherent transition radiation from a helically microbunched electron beam", E. Hemsing and J. B. Rosenzweig, *J. Applied Physics* 105, 093101 (2009)

[87] "Observation of coherent terahertz edge radiation from compressed electron beams," G. Andonian, et al., *Phys. Rev. ST Accel. Beams* 12, 030701 (2009)

[88] "Measurements of High Gain and Intensity Fluctuations in a Self-Amplified, Spontaneous-Emission Free-Electron Laser", M. Hogan, *et al. Phys. Review Lett.* **80**, 289 (1998).

[89] "Measurements of Gain Larger Than $10^5$ at 12µm in a SASE FEL", M. Hogan, *et al., Physical Review Letters,* **81** 4867 (1999).

[90] "Experimental Characterization of Nonlinear Harmonic Radiation From a Visible SASE FEL at Saturation", A. Tremaine, et al., *Phys. Rev. Lett.* **88,** 204801 (2002)

[91] "Longitudinal dispersion of orbital angular momentum modes in high-gain free-electron lasers", E. Hemsing, A. Marinelli, S. Reiche, and J.B Rosenzweig, *Phys. Rev. ST Accel. Beams,* **11,** 070704 (2008)

[92] "Generating Optical Orbital Angular Momentum in a High-Gain Free-Electron Laser at the First Harmonic", E. Hemsing, A. Marinelli, and J. B. Rosenzweig *Phys. Rev. Lett.* **106,** 164803 (2011)

[93] "Experimental Evidence of Helical Microbunching of a Relativistic Electron Beam", E. Hemsing, et al., *Applied Physics Letters* **100,** 091110 (2012)

[94] "Coherent optical vortices from relativistic electron beams", E. Hemsing, M. Dunning, D. Xiang, A. Marinelli, C. Hast, T. Raubenheimer, A. Knyazik and J. B. Rosenzweig, *Nature Physics* 9, 549 (2013).

[95] J. Harrison, et al., "Fabrication Process for micromachined multi-pole electromagnets", *Journal of MEMS Letters* **23,** 505 (2014)

[96] [I. Gadjev, et al., "An inverse free electron laser acceleration-driven Compton scattering X-ray source" *Scientific Reports* **9,** 532 (2019)

[97] M. Ferrario, et al., "EuPRAXIA@SPARC_LAB Design study towards a compact FEL facility at LNF"*, Nuclear Instruments and Methods A* 909, 134 (2018)

[98] N. Majernik, S. K. Barber, J. van Tilborg, J. B. Rosenzweig, and W. P. Leemans, "Optimization of low aspect ratio, iron dominated dipole magnets" *Phys. Rev. Accel. Beams* **22,** 032401 (2019)

[99] "Single-shot Inline Phase Contrast Imaging Using an Inverse Compton X-ray Source", P. Oliva, et al., *Appl. Phys. Lett.* 97(13) (2010)

[100] "Harmonic radiation of a relativistic nonlinear inverse Compton scattering using two laser wavelengths", Y. Sakai, et al., *Phys. Rev. ST Accel Beams*. **14,** 120702 (2011)

[101] "Single shot diffraction of picosecond 8.7-keV x-ray pulses", F. H. O'Shea, et al., *Phys. Rev. ST Accel Beams* **15**, 020702 (2012)

[102] Y. Sakai, et al.,"Observation of harmonic radiation and redshifting in the nonlinear inverse-Compton scattering interaction", *Phys. Rev. ST Accel. Beams*, 18, 060702 (2015).

[103] A. Ovodenko, et al. "High duty cycle inverse Compton scattering X-ray source" *Applied Physics Letters* DOI: http://dx.doi.org/10.1063/1.4972344 (2016)

[104] Y. Sakai, et al., "Single shot, double differential spectral measurements of inverse Compton scattering in the nonlinear regime", *Phys. Rev. Accel. Beams* **20**, 060701 (2017)

[105] J. B. Rosenzweig, et al., "An Ultra-Compact X-Ray Free-Electron Laser", *New Journal of Physics* 22, 093067 (2020)

[106] "Short period, high field cryogenic undulator for extreme performance x-ray free electron lasers" F. H. O'Shea, et al., *Phys. Rev. ST-Accel. Beams* 3, 070702 (2010)

[107] N. Majernik and James Rosenzweig, "Halbach undulators using right triangular magnets" *Phys. Rev. Accel. Beams* **22**, 092401 (2019)

[108] N. Majernik, James Rosenzweig, "Design of Comb Fabricated Halbach Undulators" *Instruments* 3(4), 58; https://doi.org/10.3390/instruments3040058 (2019)

[109] Gerard Lawler, et al., "Electron Diagnostics for Extreme High Brightness Nano-Blade Field Emission Cathodes" *Instruments* 3(4), 57; https://doi.org/10.3390/instruments3040057 (2019).

[110] Joshua Mann, Gerard Lawler, James Rosenzweig, "1-D Quantum Simulations of Electron Rescattering with Metallic Nanoblades, *Instruments* 3(4), 59; https://doi.org/10.3390/instruments3040059 (2019)

[111] R. Roussel, J. Rosenzweig, "Space radiation simulation using blowout plasm wakes at the SAMURAI Lab", *Nuclear Instruments and Methods A* http://dx.doi.org/10.1016/j.nima.2016.09.061 (2016)